\documentclass[aps,prd,twocolumn,superscriptaddress,showpacs,amsmath,amssymb,nofootinbib, preprintnumbers]{revtex4-1}
\usepackage{graphicx}
\usepackage{epstopdf}
\usepackage{float}
\usepackage{color}
\usepackage[toc,page]{appendix}
\usepackage[usenames,dvipsnames]{xcolor}

\usepackage[normalem]{ulem}

\newcommand{\be}{\begin{equation}}
\newcommand{\ee}{\end{equation}}
\newcommand{\ba}{\begin{eqnarray}}
\newcommand{\ea}{\end{eqnarray}}

\newcommand{\beq}{\begin{equation}}
\newcommand{\eeq}{\end{equation}}
\newcommand{\beqa}{\begin{eqnarray}}
\newcommand{\eeqa}{\end{eqnarray}}

\begin{document}
\title{Thermodynamics of Lorentzian Taub-NUT spacetimes}


\author{Robie A. Hennigar}
\email[]{rhennigar@mun.ca}
\affiliation{Department of Mathematics and Statistics, Memorial University of Newfoundland, St. John's, Newfoundland and Labrador, A1C 5S7, Canada}

\author{David Kubiz\v n\'ak}
\email{dkubiznak@perimeterinstitute.ca}
\affiliation{Perimeter Institute, 31 Caroline St. N., Waterloo,
Ontario, N2L 2Y5, Canada}

\author{Robert B. Mann}
\email{rbmann@sciborg.uwaterloo.ca}
\affiliation{Department of Physics and Astronomy, University of Waterloo,
Waterloo, Ontario, Canada, N2L 3G1}
\affiliation{Perimeter Institute, 31 Caroline St. N., Waterloo,
Ontario, N2L 2Y5, Canada}

\date{April 9, 2019}


\begin{abstract}
The thermodynamics of the Taub-NUT solution has been predominantly studied in the Euclidean sector, upon imposing the condition for the absence of Misner strings. Such thermodynamics is quite exceptional: the periodicity of the Euclidean time is restricted and thence the NUT charge cannot be independently varied, the entropy is not equal to $1/4$ of the area, and the thermodynamic volume  can be negative. In this paper we revisit this paradigm and study the thermodynamics of the Lorentzian Taub-NUT solution,
maintaining (as recently shown relatively harmless) Misner strings.
 We argue that in order to formulate a full cohomogeneity first law where the NUT parameter can be independently varied, it is natural to introduce a new charge together with its conjugate quantity.
We consider two scenarios: one in which the entropy is given by the Lorentzian version of the Noether charge, the other in which the entropy is given by the standard Bekenstein--Hawking area law.  In both cases consistent thermodynamics with positive thermodynamic volume can be formulated.
\end{abstract}

\maketitle

The Lorentzian Taub-NUT metric \cite{taub1951empty, newman1963empty} is one of the most intriguing (vacuum) solutions of general relativity.
 Featuring two Killing horizons and no curvature singularity, it carries a peculiar type of gravitational charge, 
the NUT charge, which is in many respects analogous to the magnetic monopole. Associated with it 
is a Misner string singularity on the polar axis 
(sometimes interpreted a singular source of angular momentum \cite{manko2005physical}) 
and the existence of spacetime regions with closed timelike curves in its vicinity.

To avoid these issues, 
Misner suggested rendering the string unobservable (similar to the Dirac string) by imposing the periodicity of the time coordinate \cite{misner1963flatter}. However, this not only leads to the existence of {closed} timelike curves everywhere, but also makes the maximal extension of the spacetime problematic, e.g. \cite{misner1963flatter, hawking1973large,hajicek1971causality}. On the other hand, when the time periodicity condition is abandoned, the Kruskal extension through both horizons is easily obtained \cite{miller1971taub}, and as recently demonstrated \cite{Clement:2015cxa, Clement:2015aka} the spacetime becomes geodesically complete and {is free from causal pathologies}
 for freely falling observers. More specifically,  for geodesics  the Misner string is completely transparent, and {(provided some restrictions are imposed on the parameters of the NUT solution---discussed below)} no closed timelike or null geodesics exist in the spacetime\footnote{It is assumed in these papers that the backreaction of non-geodesic observers (which can in principle violate causality) will modify the spacetime so that the causality is preserved.},
{thereby removing} important obstructions to recognition {of the physicality of} Lorentzian Taub-NUT spacetimes with Misner strings present.

In this paper we pursue these new exciting developments and show that, contrary to some previous doubts, e.g. \cite{Holzegel:2006gn,Kerner:2006vu},  {a consistent and reasonable} thermodynamics of the Lorentzian Taub-NUT(-Anti de Sitter) spacetime with   Misner strings present can be formulated---adding thus one more piece to the mosaic of `rehabilitation' of these spacetimes.

Let us begin our exploration by introducing the Lorentzian Taub-NUT-AdS spacetime and reviewing its basic properties.
The 
solution reads \cite{Hawking:1998ct, Clement:2015cxa}
\ba\label{j1}
ds^2&=&-f\Bigl[dt+2n(\cos\theta +\sigma) d\phi\Bigr]^2+\frac{dr^2}{f}\nonumber\\
&&\qquad \quad +(r^2+n^2)(d\theta^2+\sin^2\!\theta d\phi^2)\,,\\
f&=&\frac{r^2-2mr-n^2}{r^2+n^2}- \frac{3n^4-6n^2r^2-r^4}{l^2(r^2+n^2)}\,,
\ea
where $n$ stands for the NUT charge, $m$ for the mass parameter, and $l$ for the AdS radius,
\be
\Lambda=-\frac{3}{l^2}\,.
\ee
 Although eliminable by a `large coordinate transformation', $t\to t-2n\sigma \phi$,
the constant $\sigma$ is a physical parameter, determining the position of the Misner string(s). Namely, for $\sigma=1$ the  south  pole axis is regular, for $\sigma=-1$ the  north  one is regular, and for $\sigma=0$ both strings are `symmetrically' present.
 According to Misner \cite{misner1963flatter}, all such strings are unobservable provided the time is identified as
\be\label{Misner}
t\sim t+8\pi n\,.
\ee
In what follows we shall not impose this requirement.
As shown in \cite{Clement:2015cxa}, the spacetime is geodesically complete for any value of $\sigma$, but the requirement for the absence of closed timelike and null geodesics requires  $|\sigma|\leq 1$.


We note the characteristic behavior of NUT spacetimes, namely that the $g_{t\phi}$ component of the metric remains finite at infinity ($r\to \infty)$,
\be
g_{t\phi}\sim -2nf(\cos\theta+\sigma)\,,
\ee
with $f\sim 1$ for the asymptotically flat and $f\sim r^2/l^2$ for the asymptotically AdS Taub-NUT solutions, respectively.

The spacetime admits two Killing horizons generated by the Killing vector $\xi=\partial_t$. In what follows we concentrate on the outer horizon, located at $r_+$ given by the largest root of  $f(r_+)=0$.
  Note the proper normalization of this Killing vector at infinity, $\xi^2=-f$, and the corresponding horizon area
\be
A=4\pi (r_+^2+n^2)\,.
\ee


We now  turn to the
thermodynamics of these solutions.
Over the years, the thermodynamics of the Taub-NUT solution has been predominantly studied in the Euclidean regime (upon Wick-rotating the time  $t\to i\tau$ and the NUT parameter $n\to i\nu$) and requiring the absence of Misner {strings, e.g. } \cite{Hawking:1998ct, Chamblin:1998pz, Emparan:1999pm, Mann:1999pc, Mann:1999bt, Johnson:2014xza, Johnson:2014pwa}. Let us briefly recapitulate  this `NUTs and Bolts' approach.
In addition to the standard temperature
\be\label{TTT}
T_{\tiny{\mbox{BH}}}=\frac{f'(r_+)}{4\pi}\,,
\ee
coming from the regularity of the Euclidean solution, 
the absence of Misner strings, \eqref{Misner},  imposes a condition on the periodicity of the {Euclidean} time coordinate $\tau\sim \tau+\beta$,  $\beta=8\pi \nu$, which then leads to the following prescription for the temperature
\be\label{TT}
T_{\tiny{\mbox{S}}}= \frac{1}{8\pi \nu} \,.
\ee
Imposing equality $T_{\tiny{\mbox{S}}} = T_{\tiny{\mbox{BH}}}$ yields a nontrivial restriction on the parameters of the solution:  $\nu$ is no longer independent and becomes a function $\nu=\nu(r_+)$,
 reducing the cohomogeneity of the first law.   In the absence of other charges this simply reads
\be\label{wrong first}
\delta M= {T_{\tiny{\mbox{S}}}\delta \tilde S}+\tilde V\delta P\,,
\ee
with an entropy $\tilde S$ that is not equal to the horizon area over four, giving thus a counter-example to the Bekenstein--Hawking area law.
The latter term is present only in Anti de Sitter (AdS) spacetime, in which case the thermodynamic pressure $P$ is identified with the (negative) cosmological constant $\Lambda$, and the thermodynamic volume $\tilde V$ is the corresponding conjugate quantity \cite{Kubiznak:2016qmn},
\be
P=-\frac{\Lambda}{8\pi}\,,\quad
\tilde V=\left(\frac{\partial M}{\partial P}\right)_{\tilde{S}}\, .
\ee
Surprisingly,  for the EuclideanTaub-NUT-AdS black hole, the thermodynamic volume
\be\label{Vt}
\tilde V=\frac{4}{3}\pi r_+^3\Bigl(1-\frac{3\nu^2}{r_+^2}\Bigr)
\ee
can be negative  \cite{Johnson:2014xza, Johnson:2014pwa} (and thence also automatically violates the conjectured reverse isoperimetric inequality \cite{Cvetic:2010jb}), adding yet another `strange feature' to the thermodynamics of the Euclidean Taub-NUT solution.

We revisit this paradigm and attempt to formulate `more standard' thermodynamics for the Taub-NUT-AdS solution.
{We seek a  full cohomogeneity first law where the NUT parameter can be independently varied.}
To achieve this we
depart from   conventional wisdom by imposing the following assumptions: i) We consider the Lorentzian Taub-NUT-AdS solution instead of the Euclidean one and ii)  The (as above argued relatively harmless) Misner strings are kept present.

 These  assumptions imply that the time periodicity condition \eqref{Misner} is not imposed, and the temperature is given by the standard formula \eqref{TTT} (see \cite{Kerner:2006vu} for a quantum tunnelling derivation supporting this result):
\be\label{T}
T=\frac{1}{4\pi r_+}\Bigl(1+\frac{3(n^2+r_+^2)}{l^2}\Bigr)\,,
\ee
which is manifestly positive; the horizon is never extremal.
The NUT parameter is no longer a function of the horizon radius and can be independently varied in the first law. To maintain the full cohomogeneity, the corresponding first law thus has to have an additional term, associated with some new charge related to the NUT parameter.\footnote{The introduction of this charge seems quite natural in the light of recent developments in understanding the thermodynamics of accelerated black holes, where an extra term accounting for the string tension (causing the acceleration of the black hole)  had to be introduced \cite{Appels:2017xoe, Anabalon:2018ydc, Anabalon:2018qfv}.
}
In what follows we call this charge $N$ and its conjugate quantity $\psi$,  so that the first law takes the following form:
\be\label{first}
\delta M=T\delta S+\psi \delta N+V\delta P\,.
\ee
In order to find $N$ and $\psi$ explicitly we shall adopt two scenarios: one where the entropy
is given by the Lorentzian version of the Noether charge \cite{Garfinkle:2000ms}, and another where it is
 given by the Bekenstein--Hawking area formula. As we shall demonstrate, both scenarios lead to consistent thermodynamics with the same positive thermodynamic volume.

 To find mass and angular momentum,
we use the method of conformal completion  \cite{Ashtekar:1999jx}, which gives the following charges
\be
Q(\partial_t)=m\,,\quad Q(\partial_\phi)=3\sigma mn
\ee
respectively associated with Killing vectors $\partial_t$ and $\partial_\phi$.
Obviously, the position of Misner strings affects the angular momentum of the spacetime. In what follows we concentrate on the simplest possible case $\sigma=0$,  for which the spacetime does not contain closed timelike/null geodesics \cite{Clement:2015cxa}, the strings are `symmetrically distributed'   \cite{manko2005physical}, and the total angular momentum of the spacetime vanishes. In this case we find
\be
M=m\,,\quad J=0\,,
\ee
for the mass and angular momentum, respectively.

To proceed further, we calculate the Euclidean action
\ba\label{action}
I&=&\frac{1}{16\pi}\int_{M}d^{4}x\sqrt{g}\bigl( R+\frac{6}{\ell^{2}}\bigr)\nonumber\\
&&+ \frac{1}{8\pi}\int_{\partial M}d^{3}x\sqrt{h}\left[\mathcal{K}
-   \frac{2}{\ell} - \frac{\ell}{2}\mathcal{R}\left( h\right) \right]\,,
\ea
where  $\mathcal{K}$ and $\mathcal{R}\left( h\right)$ are respectively
the extrinsic curvature and Ricci scalar of the boundary. In this expression we have included,  apart from the Einstein--Hilbert and Gibbons--Hawking pieces, also the standard AdS counter-terms \cite{Emparan:1999pm}.
Similar to the Kerr-AdS case \cite{Caldarelli:1999xj}, in the process of calculating the action one has to not only Wick rotate the time coordinate $t\to i\tau$ but also the NUT parameter $n\to i \nu$, assume the periodicity $\tau\sim \tau+\beta$, and at the end Wick-rotate the NUT parameter back, $\nu\to -in$,
upon which we obtain the following simple result for the free energy:
\be\label{F}
F=\frac{I}{\beta}=\frac{m}{2}-\frac{1}{2l^2}(3n^2r_++r_+^3)\,.
\ee
We stress that in obtaining this result no special attention was given to the polar axis where the Misner strings are present.

Let us first turn to the `pure NUT case', setting for the moment $M=0=\Lambda$, in which case \eqref{F} yields
\be
F=0\,.
\ee
This is a rather peculiar result, as we still have a Killing horizon located at $r_+=n$ and it has a non-trivial temperature. One possibility to achieve $F=0$ is to use common wisdom and demand
\be\label{F1}
F=M-TS\,,
\ee
to infer that (contrary to the area law) we have to have $S=0$.
Another possibility is to consider positive entropy and introduce conjugates $\psi$ and $N$ in the free energy formula in a manner similar to the $\Omega J$ term in the grand-canonical ensemble
\be\label{F2}
F=M-TS-\psi N\,.
\ee
Let us look at these more closely for generic $M$ and $\Lambda$.

Demanding \eqref{F1},   \eqref{F} yields the following expression for the entropy:
\be\label{SNC}
S=S_{\tiny{\mbox{NC}}}=\frac{\pi(3r_+^4+12n^2r_+^2+r_+^2l^2-n^2l^2-3n^4)}{3n^2+l^2+3r_+^2}\,,
\ee
which is the Lorentzian version of {the  entropy obtained by Noether charge methods} \cite{Garfinkle:2000ms}.
 We then find that the following quantities:
\be\label{psi-N1}
\psi=-\frac{n(l^2+3n^2-3r_+^2)}{2(3n^2+l^2+3r_+^2)}\,, \quad N=\frac{n}{r_+}+\frac{3n(n^2+r_+^2)}{r_+l^2}\,,
\ee
together with the volume
\be\label{PV}
V=\frac{4}{3}\pi r_+^3\Bigl(1+\frac{3n^2}{r_+^2}\Bigr)\,,
\ee
satisfy the first law \eqref{first} and the Smarr formula
\be\label{Smarr1}
M=2(TS-VP)\,,
\ee
with the latter consistent with the first law by the dimensional scaling argument  \cite{Kastor:2009wy}.
Note that the NUT potential $\Psi$ and charge $N$ both vanish as $n\to 0$.

The second possibility \eqref{F2} is not unique, unless we specify the form of the entropy. In what follows we concentrate on the `natural choice'
and   require the entropy to be given by the Bekenstein--Hawking area law:
\be\label{SBH}
S=S_{\tiny{\mbox{BH}}}= \frac{A}{4}=\pi (r_+^2+n^2)\,.
\ee
The quantities $\psi$ and $N$ are then uniquelly determined and read
\be\label{quantities}
\psi=\frac{1}{8\pi n}\,,\quad N=-\frac{4\pi n^3}{r_+}+\frac{12\pi r_+n^3}{l^2}\Bigl(1-\frac{n^2}{r_+^2}\Bigr)\,.
\ee
Together with the same $P$ and $V$, \eqref{PV}, they satisfy the first law \eqref{first} and the modified Smarr formula:
\be\label{Smarr2}
M=2(TS-VP+\psi N)\,,
\ee
with the two consistent by the dimensional scaling argument \cite{Kastor:2009wy}.

Note that both sets of thermodynamic quantities have a smooth asymptotically flat limit, $\Lambda\to 0$, yielding thus a consistent thermodynamics of asymptotically flat Taub-NUT spacetimes.  Furthermore, the  thermodynamic volume $V$, \eqref{PV}, is manifestly positive\footnote{Note that the possible negativity of \eqref{Vt} is simply an artifact of the the Wick rotation of the NUT charge, c.f. \eqref{Vt} and \eqref{PV}.
} and the same for each case.
Moreover, the isoperimetric ratio
(taking $\omega_2=4\pi$) reads
\be
\mathcal{R}=\left( \frac{3{V}}{\omega _{2}}\right)^{\frac{1}{3}}
\left(\frac{\omega _{2}}{{A}}\right) ^{\frac{1}{2}}=\frac{(1+3q^2)^{1/3}}{\sqrt{1+q^2}}\,,\quad q=\frac{n}{r_+}\,.
\ee
The reverse isoperimetric inequality \cite{Cvetic:2010jb} requires $\mathcal{R}\geq 1$, which is true for $q\in [0,\sqrt{3+2\sqrt{3}}\approx 2.54]$. It is easy to show {that $q \leq \sqrt{1+2/\sqrt{3}} \approx 1.47$ for horizons characterized by positive $M$,  and so} the reverse isoperimetric inequality holds for the Lorentzian Taub-NUT-AdS spacetimes.\footnote{ The horizon structure changes upon the Wick rotation $n\to i\nu$. Although in the Lorentzian case horizons can exist even for $\quad M<0$,~we have concentrated here on positive mass solutions. }

Our results imply that  Lorentzian Taub-NUT spacetimes, with   Misner strings present, are much less pathological than previously expected---in addition to admitting a Kruskal extension through both Killing horizons, being geodesically complete and not violating causality for geodesic observers
(despite containing regions with closed timelike curves), they also admit a consistent thermodynamic formulation.   The thermodynamic first law \eqref{first} and the associated  Smarr formulae \eqref{Smarr1} and \eqref{Smarr2}  can be consistently formulated in two distinct scenarios,
opening a window for further physical interpretation of these solutions and their potential astrophysical applications (for example,   possible NUT signatures in   microlensing data have
been considered \cite{lynden1998classical, bogdanov2008search}).

The key ingredient for formulating the `Lorentzian thermodynamics' of the Taub-NUT solution proposed in this paper is to  maintain Misner strings, relaxing the time periodicity condition \eqref{Misner}. Consequently (and contrary to the Euclidean case) the horizon radius and the NUT parameter
are treated as independent and the full cohomogeneity first law, with a new conjugate pair $\psi-N$, can be written down.
Two possible scenarios are viable, and we conclude by commenting on each.

By computing entropy via the Noether charge method  \cite{Garfinkle:2000ms}, we find that it is consistent with the expected expression \eqref{F1} for the free energy.   The (AdS) Taub-NUT black hole, regarded as a single system, has an entropy smaller than that given by the area law, and the temperature is given by \eqref{T}.  No attempt is made to impose identification with \eqref{TT}; this formula is obtained by demanding absence of the string in the Euclidean section and upon analytic continuation, yields
an imaginary value for the temperature.   Indeed, given
the benign nature of the Misner string \cite{Clement:2015cxa, Clement:2015aka}, this additional constraint is not necessary.
However the set of variables \eqref{SNC}, \eqref{psi-N1}, and \eqref{PV}
 has the peculiar feature that  if $M=0$, $S=0$ (implying likewise that $F=0$), but the temperature \eqref{T} remains non-vanishing.  In this case the spacetime has
finite temperature but has vanishing entropy---a situation unprecedented in statistical physics.  There is also some ambiguity in determining the conjugate pair $\psi-N$ since it does not appear in either the action or  the Smarr relation; however this ambiguity can be fixed by demanding that both vanish {linearly} as
$n\to 0$.

A  perhaps more promising alternative is to require that  the entropy  of the system is given by the Bekenstein--Hawking area law for the outer Killing horizon. That is, contrary to the conventional wisdom applied in the
Euclidean section (e.g. \cite{Carlip:1999cy, Emparan:1999pm, Mann:1999pc, Garfinkle:2000ms}) we have not assigned any entropy to the Misner string on the axis.  The new conjugate pair $\psi-N$ is determined by requiring \eqref{F2} to hold in a manner consistent
with the first law, with the peculiar feature that $\psi$ diverges as $n\to 0$.   This situation is analogous to the thermodynamics of accelerated black holes, where the area law still applies despite the presence of cosmic strings on the axis \cite{Appels:2017xoe, Anabalon:2018ydc, Anabalon:2018qfv}.

 In the Euclidean case, it was argued \cite{Carlip:1999cy} that Misner strings located on the north/south axis correspond to a Killing horizon generated by $\partial_\tau\pm \frac{1}{2\nu}\partial_\phi$ which yields the temperature \eqref{TT} and has associated with it (negative) degrees of freedom.
This is rather questionable in the Lorentzian {case} where the spacetime is geodesically complete and such a horizon does not really `hide any information', suggesting that \eqref{SBH} may perhaps be a better choice than \eqref{SNC}.
Note, however, that in this case
one might regard $\psi$ and $N$ in \eqref{quantities}
as the respective Lorentzian temperature and entropy of the Misner string. If such an interpretation is correct, one would have
(similar to de Sitter black hole spacetimes, e.g. \cite{Dolan:2013ft}) a system out of thermodynamic equilibrium unless $\psi=1/(8\pi n)$ were set equal to the temperature $T$
in \eqref{T}. This would reduce the cohomogeneity of the system, rendering thermodynamics qualitatively similar
(but quantitatively different) to the Euclidean case \cite{Carlip:1999cy, Emparan:1999pm, Mann:1999pc, Garfinkle:2000ms}.
{Let us finally note that a proposal similar to \eqref{SBH} and \eqref{quantities} was recently advocated in \cite{Bueno:2018uoy} in the context of Euclidean Taub solutions with toroidal base spaces. Such solutions do not possess Misner strings, but consistent thermodynamics requires the introduction of a new potential $\psi \propto 1/\nu$.}

Our proposed approach to thermodynamics of the Taub-NUT spacetimes raises interesting questions regarding the nature of gravitational thermodynamics.
What is the physical meaning of the new charge $N$ and its conjugate quantity $\psi$?  Can consistent thermodynamics be formulated also for the Taub-NUT solutions with additional charges and/or rotation, and their generalizations to higher dimensions and beyond Einstein gravity? Can we find any interesting  thermodynamic phase transitions,  similar to what is observed for the AdS black hole spacetimes \cite{Kubiznak:2016qmn}? Even more generally, can the obtained thermodynamics be used to understand the physical sources of the Taub-NUT solutions? And finally, taking into account all the obstructions removed, can these spacetimes be astrophysically realized?

\section*{Acknowledgements}
We are grateful to Pablo Cano and Ruth Gregory for discussions.  This work was supported in part by the Natural Sciences and Engineering Research Council of Canada.
R.H. is grateful to the Banting
Postdoctoral Fellowship programme.
D.K.\ acknowledges the Perimeter Institute for Theoretical Physics  for their support. Research at Perimeter Institute is supported by the Government of Canada through the Department of Innovation, Science and Economic Development Canada and by the Province of Ontario through the Ministry of Research, Innovation and Science.


\providecommand{\href}[2]{#2}\begingroup\raggedright\endgroup

\end{document}